\documentclass[aps,twocolumn,pra,superscriptaddress,amsmath,showpacs,tightenlines]{revtex4}
\usepackage{epsfig,graphicx,times}
\usepackage{amstext}
\usepackage{amsmath}            
\usepackage{amssymb}            
\usepackage{graphicx}           
\usepackage{latexsym}
\usepackage{bm}
\usepackage[colorlinks,citecolor=blue, linkcolor=blue,hyperindex,CJKbookmarks,dvipdfm]{hyperref}

\begin{document}

\title{Tunable photon statistics in weakly nonlinear photonic molecules}
\author{Xun-Wei Xu}
\affiliation{Beijing Computational Science Research Center, Beijing 100084, China}
\author{Yong Li}
\email{liyong@csrc.ac.cn}
\affiliation{Beijing Computational Science Research Center, Beijing 100084, China}
\affiliation{Synergetic Innovation Center of Quantum Information and Quantum Physics,
University of Science and Technology of China, Hefei, Anhui 230026, China}
\date{\today }

\begin{abstract}
In recent studies [Liew \emph{et al.}, Phys. Rev. Lett.~\textbf{104}, 183601
(2010); Bamba \emph{et al.}, Phys. Rev. A~\textbf{83}, 021802(R) (2011)],
due to destructive interference between different paths for two-photon
excitation, strong photon antibunching can be obtained in a photonic
molecule consisting of two coupled cavity modes with weak Kerr nonlinearity
when one of the cavity modes is driven resonantly. Here, we study the photon
statistics in a nonlinear photonic molecule with both the two cavity modes being driven
coherently. We show that the statistical properties of the photons can be
controlled by regulating the coupling constant between the cavity modes, the strength ratio and the relative phase between the driving fields. The photonic molecules with two driven modes can be used to generate tunable single-photon sources or controlled photonic quantum gates with weak Kerr nonlinearity.
\end{abstract}

\maketitle


\section{Introduction}

Single-photon sources play an important role in quantum cryptography and
quantum communication~\cite{LounisRPP05,ScaraniRMP09}. Perfect single-photon
sources emit photons one by one, i.e. the photons exhibit antibunching effect. Photon blockade that the excitation of a first photon blocks the transport
of a second photon for the nonlinear medium in the cavity is one of
the mechanisms to create antibunching photons~\cite{ImamogluPRL97}. For convenient photon blockade, strong nonlinear interaction is one of the necessary requisites~\cite{ImamogluPRL97}. A sequence of experimental groups observed the photon blockade in different systems, such as an optical cavity with one
trapped atom~\cite{BirnbaumNat05, DayanSci08,DubinNP10}, a quantum dot in a photonic
crystal~\cite{FaraonNP08}, circuit quantum electrodynamics (cQED)
system~\cite{LangPRL11,HoffmanPRL11,LiuPRA14}, etc.

Recently, a new mechanism called unconventional photon blockade (UPB) was
found by Liew and Savona~\cite{LiewPRL10}. They showed that strong photon
antibunching can be obtained in two coupled cavities (photonic molecule~\cite{BayerPRL98,RakovichLPR10}) with weak nonlinearities in the cavities~\cite{LiewPRL10}. This surprisingly strong antibunching was originated from the destructive quantum interference effect in the nonlinear photonic molecule~\cite{BambaPRA11}. This mechanism is universal and many different nonlinear systems are proposed to realize the UPB, including bimodal optical cavity with a quantum dot~\cite{MajumdarPRL12,ZhangPRA14}, coupled optomechanical systems~\cite{XuJPB13,SavonaARX13}, or coupled single-mode cavities with second- or third-order nonlinearity~\cite{FerrettiNJP13,FlayacPRA13,GeracePRA14}.

The optimal conditions for strong antibunching in a photonic molecule when
one of the cavity modes is driven resonantly have been given in Refs.~\cite{LiewPRL10,BambaPRA11}, where the optimal detuning is only dependent on the dissipation rates of the cavity modes in the strong coupling condition. However,
for general optical cavities, it is difficult to adjust the dissipation rates
of the cavity modes. So it should be interesting that how to create
tunable single-photon sources where the optimal conditions for strong
antibunching are related to some controllable parameters in
experiments. Kyriienko \emph{et al.} proposed a tunable single-photon
emission from dipolaritons by embedding a double quantum well in a
micropillar optical cavity~\cite{KyriienkoarX14}. They found that the
equal-time second-order correlation function of the dipolaritons in the
hybrid system can be tuned by using an electric field applied to the structure,
or changing the frequency of the pump. In our previous paper, we studied the
photon statistics of symmetric and antisymmetric modes in a photonic
molecule consisting of two coupled cavities with weak nonlinearity, and found that the optimal frequency detuning for strong photon antibunching of the symmetric and antisymmetric modes is linearly dependent on the coupling constant between the cavity modes in the photonic molecule,
which provides us another way to generate tunable single-photon sources~\cite{XuarX14} since the coupling between the cavity modes can be controlled experimentally, e.g., by changing the distance between the coupled cavities~\cite{GrudininPRL10,PengOL12,PengNP14a,ChangNP14,PengNP14b}.

In this paper, in order to make the conditions for strong photon antibunching of local cavity modes in a photonic molecule easily tunable in experiments, we extend the works of Refs.~\cite{LiewPRL10,BambaPRA11} for both the two cavity modes being driven coherently~\cite{ZhangPRA14}. Our calculations show that the optimal conditions for strong photon antibunching of local cavity modes are dependent not only on the coupling constant between the cavity modes, but also on the strength ratio and the relative phase between the two driving
fields. So two more tunable parameters (compared with Ref.~\cite{XuarX14}), i.e., the strength ratio and the relative phase between the driving fields, can be used for obtaining tunable single-photon
sources when both the two cavity modes are driven coherently. What is more, strong bunching effect can also be obtained in some special conditions. Thus the photonic molecules with two driven modes can be used to generate controlled photonic quantum gates.

The paper is organized as follows: In Sec.~II, we derive the optimal
antibunching conditions analytically for the two cavity modes in the photonic molecule system are both driven weakly. In Sec.~III, the dependence of the
statistic properties of the photons on the parameters in the photonic molecule is investigated via the second-order correlation functions
numerically. Finally, we summary our paper in Sec.~IV.

\section{Optimal antibunching conditions}

We consider a photonic molecule consisting of two driven nonlinear cavity modes (A and B), which can be achieved experimentally in the system of two coupled whispering-gallery-mode optical resonators~\cite{GrudininPRL10,PengOL12,PengNP14a,ChangNP14,PengNP14b}. In a frame rotating at the identical frequency of the two driving fields $\omega _{d}^{a}=\omega _{d}^{b}=\omega _{d}$, the Hamiltonian for the system reads ($\hbar =1$)
\begin{eqnarray}
H &=&\Delta _{a}a^{\dag }a+\Delta _{b}b^{\dag }b+J\left( ab^{\dag }+a^{\dag
}b\right)   \notag \\
&&+U_{a}a^{\dag }a^{\dag }aa+U_{b}b^{\dag }b^{\dag }bb   \notag\\
&&+\left( \varepsilon _{a}e^{i\phi _{a}}a^{\dag }+\varepsilon _{b}e^{i\phi
_{b}}b^{\dag }+\rm{H.c.}\right) ,  \label{eq:1}
\end{eqnarray}%
where $a$ ($b$) is the bosonic operator eliminating a photon in cavity mode A (B) with frequency $\omega _{a}$ ($\omega _{b}$), $U_{a}$ ($U_{b}$) is the Kerr nonlinear interaction strength, and $J$ is the real coupling constant between the cavity modes. $\varepsilon _{a}$ ($\varepsilon _{b}$) and $\phi _{a}$ ($\phi _{b}$) are the real strength and phase of the external driving fields with frequency $\omega _{d}$. $\Delta _{a}=\omega
_{a}-\omega _{d}$ ($\Delta _{b}=\omega _{b}-\omega _{d}$) is the frequency
detuning between the cavity mode A (B) and the related driving field. In the following, we will
consider the dissipations for both cavity modes with dissipation rates $\kappa _{a}$ and $\kappa _{b}$. Even though the best combination of the detunings and dissipations can
lead to further optimal photon antibunching~\cite{FerrettiNJP13}, for
distinct physical picture and brief results, here we assume that $\kappa
_{a}=\kappa _{b}=\kappa $ and $\Delta _{a}=\Delta _{b}=\Delta $.

As shown in Ref.~\cite{BambaPRA11}, in the weak driving and strong coupling
conditions $J\gg \kappa \gg \varepsilon _{a}$ and without driving cavity mode B ($\varepsilon _{b}=0$), the optimal conditions for photons in cavity mode A
exhibiting strong antibunching are given by
\begin{eqnarray}
\Delta _{\mathrm{opt}} &\approx &\pm \frac{\kappa }{2\sqrt{3}},  \label{eq:2}
\\
U_{\mathrm{opt}} &\approx &\pm \frac{2}{3\sqrt{3}}\frac{\kappa ^{3}}{J^{2}},
\label{eq:3}
\end{eqnarray}%
where the optimal nonlinear interaction strength is required only in cavity mode B.

Following the method given in Ref.~\cite{BambaPRA11}, we will derive the
optimal conditions for the case that both the two cavity modes are driven coherently. In the weak driving condition $\varepsilon _{a,b}\ll \kappa $, we can expand the wave function on a Fock-state basis truncated to the two-photon
manifold with the ansatz
\begin{eqnarray}
\left\vert \psi \right\rangle &=&C_{00}\left\vert 0,0\right\rangle
+C_{10}\left\vert 1,0\right\rangle +C_{01}\left\vert 0,1\right\rangle  \notag
\\
&&+C_{20}\left\vert 2,0\right\rangle +C_{11}\left\vert 1,1\right\rangle
+C_{02}\left\vert 0,2\right\rangle .  \label{eq:4}
\end{eqnarray}%
Here, $\left\vert n_{a},n_{b}\right\rangle $ represents the Fock state with $%
n_{a}$ photons in mode A and $n_{b}$ photons in mode B. By substituting the
wave function [Eq.~(\ref{eq:4})] and Hamiltonian [Eq.~(\ref{eq:1})] into the
Schr\"{o}dinger's equation, we get the dynamic equations for the coefficients $C_{n_{a}n_{b}}$
\begin{widetext}
\begin{eqnarray}
i\frac{\partial }{\partial t}C_{00} &=&\varepsilon _{a}e^{-i\phi
_{a}}C_{10}+\varepsilon _{b}e^{-i\phi _{b}}C_{01},  \label{eq:6} \\
i\frac{\partial }{\partial t}C_{10} &=&\left( \Delta -i\frac{\kappa }{2}%
\right) C_{10}+JC_{01}+\varepsilon _{a}e^{i\phi _{a}}C_{00}+\varepsilon
_{b}e^{-i\phi _{b}}C_{11}+\sqrt{2}\varepsilon _{a}e^{-i\phi _{a}}C_{20},
\label{eq:7} \\
i\frac{\partial }{\partial t}C_{01} &=&\left( \Delta -i\frac{\kappa }{2}%
\right) C_{01}+JC_{10}+\varepsilon _{b}e^{i\phi _{b}}C_{00}+\varepsilon
_{a}e^{-i\phi _{a}}C_{11}+\sqrt{2}\varepsilon _{b}e^{-i\phi _{b}}C_{02},
\label{eq:8} \\
i\frac{\partial }{\partial t}C_{11} &=&\left( 2\Delta -i\kappa \right)
C_{11}+\sqrt{2}J\left( C_{20}+C_{02}\right) +\varepsilon _{b}e^{i\phi
_{b}}C_{10}+\varepsilon _{a}e^{i\phi _{a}}C_{01},  \label{eq:9} \\
i\frac{\partial }{\partial t}C_{20} &=&\left( 2\Delta +2U_{a}-i\kappa
\right) C_{20}+\sqrt{2}JC_{11}+\sqrt{2}\varepsilon _{a}e^{i\phi _{a}}C_{10},
\label{eq:10} \\
i\frac{\partial }{\partial t}C_{02} &=&\left( 2\Delta +2U_{b}-i\kappa
\right) C_{02}+\sqrt{2}JC_{11}+\sqrt{2}\varepsilon _{b}e^{i\phi _{b}}C_{01}.
\label{eq:11}
\end{eqnarray}

Under the weak driving condition $\varepsilon _{a,b}\ll \kappa $, we have $|C_{00}|\gg \{|C_{10}|,|C_{01}|\} \gg \{|C_{20}|,|C_{11}|,|C_{02}|\}$. In the
steady state, $\partial C_{n_{a}n_{b}}/\partial t=0$, the equations for the
coefficients of one-photon states are given approximately as
\begin{eqnarray}
\left( \Delta -i\frac{\kappa }{2}\right) C_{10}+JC_{01} &=&-\varepsilon
_{a}e^{i\phi _{a}}C_{00},  \label{eq:12} \\
JC_{10}+\left( \Delta -i\frac{\kappa }{2}\right) C_{01} &=&-\varepsilon
_{b}e^{i\phi _{b}}C_{00},  \label{eq:13}
\end{eqnarray}%
and the equations for the coefficients of two-photon states are given as
\begin{eqnarray}
0 &=&\left( 2\Delta +2U_{a}-i\kappa \right) C_{20}+\sqrt{2}JC_{11}+\sqrt{2}%
\varepsilon _{a}e^{i\phi _{a}}C_{10},  \label{eq:14} \\
0 &=&\left( 2\Delta +2U_{b}-i\kappa \right) C_{02}+\sqrt{2}JC_{11}+\sqrt{2}%
\varepsilon _{b}e^{i\phi _{b}}C_{01},  \label{eq:15} \\
0 &=&\left( 2\Delta -i\kappa \right) C_{11}+\sqrt{2}J\left(
C_{20}+C_{02}\right) +\varepsilon _{b}e^{i\phi _{b}}C_{10}+\varepsilon
_{a}e^{i\phi _{a}}C_{01}.  \label{eq:16}
\end{eqnarray}%
From Eqs.~(\ref{eq:12}) and (\ref{eq:13}), $C_{10}$\ and $C_{01}$ are obtained
\begin{eqnarray}
C_{10} &=&\frac{\left[ \varepsilon _{b}e^{i\phi _{b}}J-\varepsilon
_{a}e^{i\phi _{a}}\left( \Delta -i\frac{\kappa }{2}\right) \right] }{\left(
\Delta -i\frac{\kappa }{2}\right) ^{2}-J^{2}}C_{00},  \label{eq:17} \\
C_{01} &=&\frac{\left[ \varepsilon _{a}e^{i\phi _{a}}J-\varepsilon
_{b}e^{i\phi _{b}}\left( \Delta -i\frac{\kappa }{2}\right) \right] }{\left(
\Delta -i\frac{\kappa }{2}\right) ^{2}-J^{2}}C_{00}.  \label{eq:18}
\end{eqnarray}%
By substituting the above expressions of $C_{10}$\ and $C_{01}$ into Eqs.~(\ref{eq:14})-(\ref{eq:16}), we get
\begin{eqnarray}
0 &=&\left( 2\Delta +2U_{a}-i\kappa \right) C_{20}+\sqrt{2}JC_{11}+\frac{%
\sqrt{2}\left[ J-\eta e^{i\phi }\left( \Delta -\frac{i\kappa }{2}\right) %
\right] }{\left( \Delta -i\frac{\kappa }{2}\right) ^{2}-J^{2}}\widetilde{C}%
_{00},  \label{eq:19} \\
0 &=&\left( 2\Delta +2U_{b}-i\kappa \right) C_{02}+\sqrt{2}JC_{11}+\frac{%
\sqrt{2}\left[ J-\eta ^{-1}e^{-i\phi }\left( \Delta -\frac{i\kappa }{2}%
\right) \right] }{\left( \Delta -i\frac{\kappa }{2}\right) ^{2}-J^{2}}%
\widetilde{C}_{00},  \label{eq:20} \\
0 &=&\left( 2\Delta -i\kappa \right) C_{11}+\sqrt{2}J\left(
C_{20}+C_{02}\right) +\frac{\left( \eta ^{-1}e^{-i\phi }+\eta e^{i\phi
}\right) J-\left( 2\Delta -i\kappa \right) }{\left( \Delta -i\frac{\kappa }{2%
}\right) ^{2}-J^{2}}\widetilde{C}_{00},  \label{eq:21}
\end{eqnarray}%
\end{widetext}
where $\widetilde{C}_{00}\equiv C_{00}\varepsilon _{a}\varepsilon _{b}e^{i\left(
\phi _{a}+\phi _{b}\right) }$, and the coefficients $\widetilde{C}_{00}$, $C_{11}$, $C_{20}$ and $C_{02}$ are dependent on the strength ratio $\eta \equiv\varepsilon _{a}/\varepsilon _{b}$ (or $\eta^{-1} \equiv\varepsilon _{b}/\varepsilon _{a}$) and the relative phase $\phi \equiv\phi _{a}-\phi _{b}$ between the driving fields.

As the system is symmetric for cavity modes A and B, in order to avoid the redundancy, we will focus on the photon statistics of cavity mode A. The equal-time
second-order correlation function of the photons in mode A, $g_{a}^{\left(
2\right) }\left( 0\right)\equiv
\left\langle a^{\dag }a^{\dag }aa\right\rangle /\left\langle a^{\dag }a\right\rangle ^2$, can be calculated by solving the master equations numerically~\cite{VergerPRB06,LiewPRL10,BambaPRA11}. In the weak driving condition, $|C_{\left( n_{a}+1\right) n_{b}}|^{2}\ll
|C_{n_{a}n_{b}}|^{2}$, $|C_{n_{a}\left( n_{b}+1\right)}|^{2}\ll
|C_{n_{a}n_{b}}|^{2}$, $g_{a}^{\left( 2\right) }\left( 0\right) $ can be
given approximately by~\cite{XuPRA13}
\begin{equation}
g_{a}^{\left( 2\right) }\left( 0\right) \approx \frac{2|C_{20}|^{2}}{|C_{10}|^{2}}. \label{eq:21b}
\end{equation}%
The conditions for $g_{a}^{\left( 2\right) }\left( 0\right) \ll 1$ are
equivalent to that for $C_{20} \approx 0$ in Eqs.~(\ref{eq:19})-(\ref{eq:21}). By setting $C_{20}=0$ in Eqs.~(\ref{eq:19})-(\ref{eq:21}), the
condition for $\widetilde{C}_{00}$, $C_{11}$ and $C_{02}$ having non-trivial
solutions is that the determinant of the corresponding coefficient matrices equals to zero. As $U_{a}$ does not appear in the coefficients matrices after setting $C_{20}=0$, the optimal nonlinear interaction strength is required only in mode B~\cite{BambaPRA11}. However, the general optimal antibunching conditions (including the strength ratio $\eta$ and the relative phase $\phi$ between the driving fields) are too cumbersome and not present here.

In the special case of $\phi =0$, to make the imaginary and real parts of the determinant of the coefficient matrices for Eqs.~(\ref{eq:19})-(\ref{eq:21}) equal to zero, we get the equations for the optimal parameters as
\begin{eqnarray}
0 &=&16J\Delta ^{2}-4J\kappa ^{2}+6\Delta \kappa ^{2}\eta -8\Delta ^{3}\eta
\notag \\
&&-8\Delta J^{2}\eta ^{-1}+16J\Delta U-4J^{2}\eta ^{-1}U  \notag \\
&&-4J^{2}\eta U-8\Delta ^{2}\eta U+2\kappa^{2}\eta U,  \label{eq:22} \\
0 &=&4J^{2}\kappa \eta ^{-1}+12\kappa \Delta ^{2}\eta -\kappa ^{3}\eta
\notag \\
&&-16J\kappa \Delta+8\kappa \Delta \eta U-8J\kappa U.  \label{eq:23}
\end{eqnarray}%
In the strong coupling condition $J\gg \kappa $ and also $(J\Delta) \gg \kappa^{2} $, the parameters should satisfy the following conditions
\begin{eqnarray}
\Delta _{\mathrm{opt}} &\approx &J/\eta,  \label{eq:24} \\
U_{\mathrm{opt}} &\approx &\frac{\kappa ^{2}}{2J}\frac{\eta }{\left( \eta
^{2}-1\right) }.  \label{eq:25}
\end{eqnarray}%
Eqs.~(\ref{eq:24}) and (\ref{eq:25}) are the optimal conditions for $%
g_{a}^{\left( 2\right) }\left( 0\right) \ll 1$ when $\phi =0$. From Eqs.~(%
\ref{eq:24}) and (\ref{eq:25}), the optimal conditions for strong antibunching
are dependent on the coupling constant $J$ and the strength ratio between the
driving fields $\eta$. These imply that we can control the statistic
properties of the photons by tuning the coupling constant between the
cavity modes (as shown in Ref.~\cite{XuarX14}) or the strength ratio between the two driving fields.

Let us do some discussions about Eqs.~(\ref{eq:24}) and (\ref{eq:25}). First of
all, in order to make sure that the strong antibunching occurs in mode A in
the weak nonlinear regime, i.e. $U_{\mathrm{opt}}/\kappa < 1$, the strength
ratio $\eta$ should be larger than one, $\eta>1$, i.e. $\varepsilon
_{a}>\varepsilon _{b}$. Second, Eqs.~(\ref{eq:24}) and (\ref{eq:25}) are
applicable only in the case for $\eta \ll (J/\kappa)^2$. Because as $\eta
\rightarrow +\infty$, we have $\Delta _{\mathrm{opt}} \rightarrow 0$ and $U_{
\mathrm{opt}} \rightarrow 0$ from Eqs.~(\ref{eq:24}) and (\ref{eq:25}), which do
not agree with the results given in Eqs.~(\ref{eq:2}) and (\ref{eq:3}) for the
case of $\varepsilon _{b} =0$. In the next section, we will analyse
numerically about the regime that Eqs.~(\ref{eq:24}) and (\ref{eq:25}) are
applicable.

\section{Numerical results}

\begin{figure}[tbp]
\includegraphics[bb=88 278 485 551, width=4.2 cm, clip]{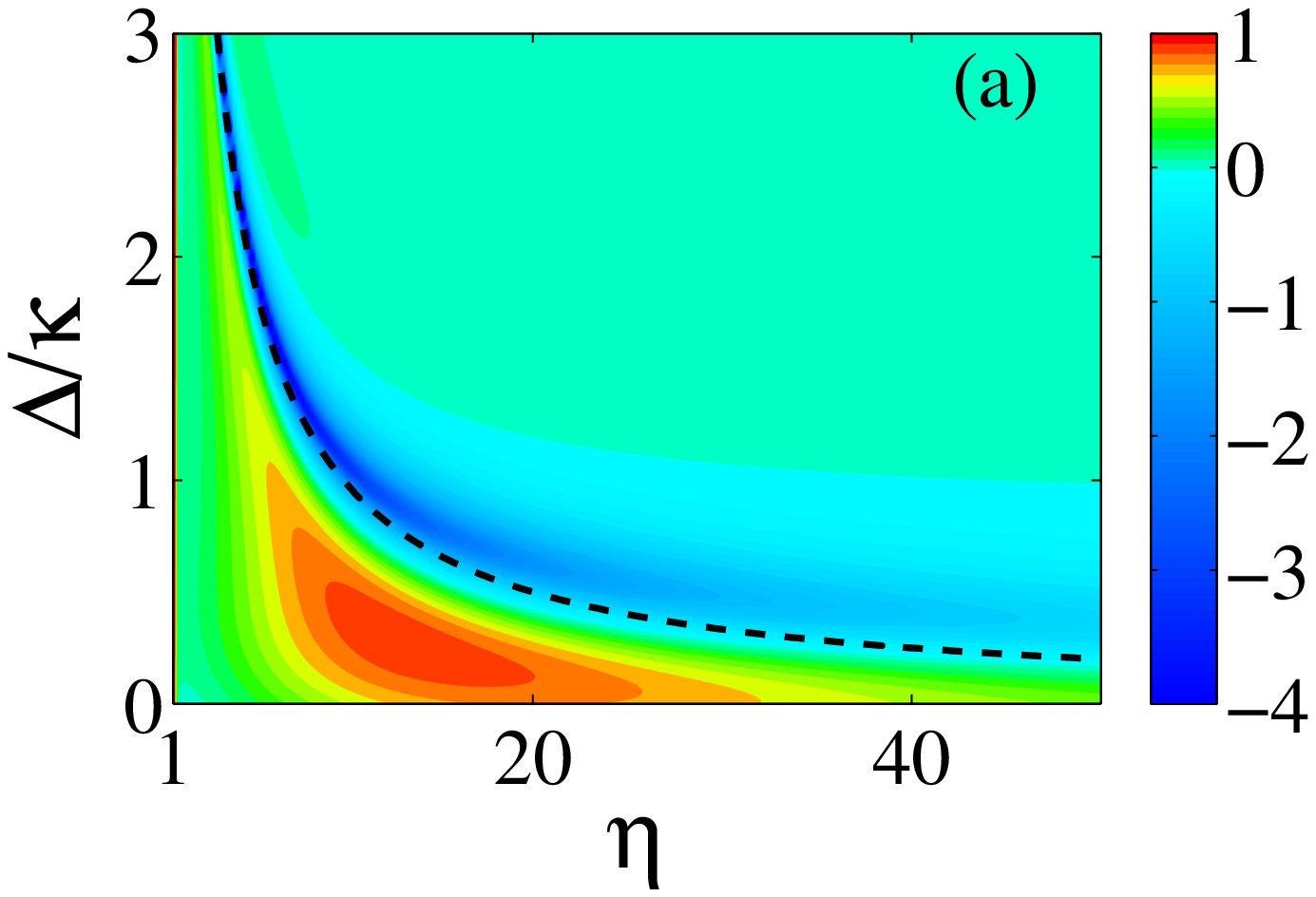} %
\includegraphics[bb=88 278 485 551, width=4.2 cm, clip]{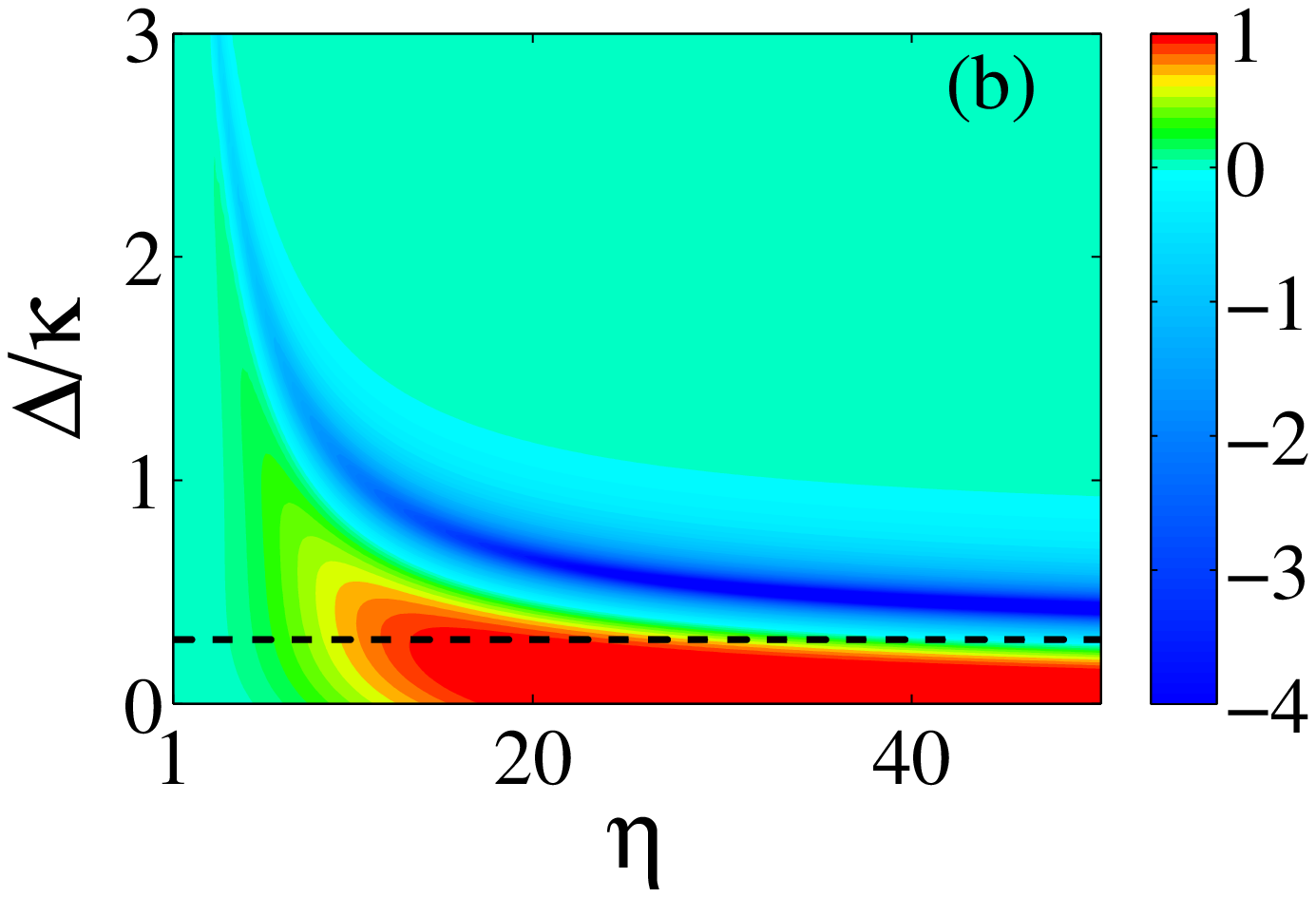}
\caption{(Color online) Logarithmic plot (of base $10$) of the equal-time
second-order correlation function $g_{a}^{\left( 2\right) }\left( 0\right)$
as a function of the strength ratio between the driving fields $\protect\eta$
and the detuning $\Delta/\kappa$ with the nonlinear interaction
strength $U$ satisfying Eq.~(\ref{eq:25}) in (a) or  satisfying Eq.~(\ref{eq:3}) in (b).
The dash line in (a) [(b)] corresponds to the detuning $\Delta/\kappa$ satisfying Eq.~(\ref{eq:24}) [Eq.~(\ref{eq:2})].
The parameters are $J= 10 \protect\kappa$ and $\protect\phi=0$.}
\label{fig1}
\end{figure}

Next, we will calculate the equal-time second-order correlation function $%
g_{a}^{\left( 2\right) }\left( 0\right)$ by solving the master equation
numerically within a truncated Fock space~\cite%
{VergerPRB06,LiewPRL10,BambaPRA11}. Here both the two external driving fields are weak~\cite{LiuPRA14} and we set the strength of the driving fields to mode A and B as $\varepsilon_{b}<\varepsilon_{a}=0.01\kappa$.

In Fig.~\ref{fig1}, we show the equal-time second-order correlation
function $g_{a}^{\left( 2\right) }\left( 0\right) $ as a function of the
strength ratio between the driving fields $\eta$ and the detuning $%
\Delta/\kappa$ with the nonlinear interaction strength $U$ satisfying Eq.~(\ref{eq:25}) in (a) or with $U$ satisfying Eq.~(\ref{eq:3}) in (b). From Fig.~\ref{fig1}(a), as the strength ratio $\eta$ is not too large, e.g. $\eta<10$, the antibunching effect is strong; with the increase of $\eta$, the strength for antibunching becomes weaker. This shows that the
optimal condition for antibunching in Eq.~(\ref{eq:25}) is not applicable as
$\eta$ is too large. In this case of large $\eta$, the optimal condition Eq.~(\ref{eq:3}) is suitable, as shown in Fig.~\ref{fig1}(b).

\begin{figure}[tbp]
\includegraphics[bb=88 278 485 551, width=4.2 cm, clip]{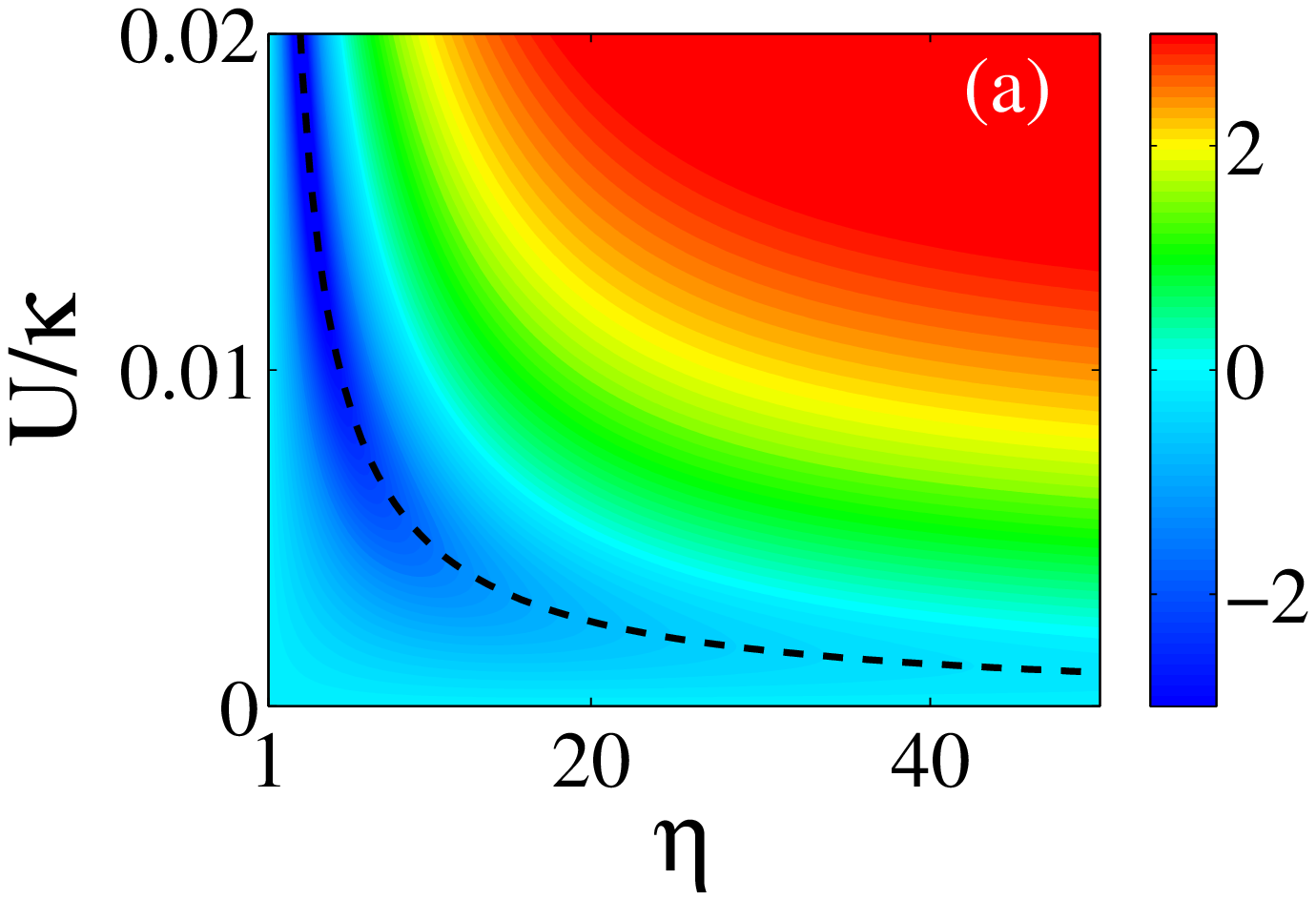} %
\includegraphics[bb=88 278 485 551, width=4.2 cm, clip]{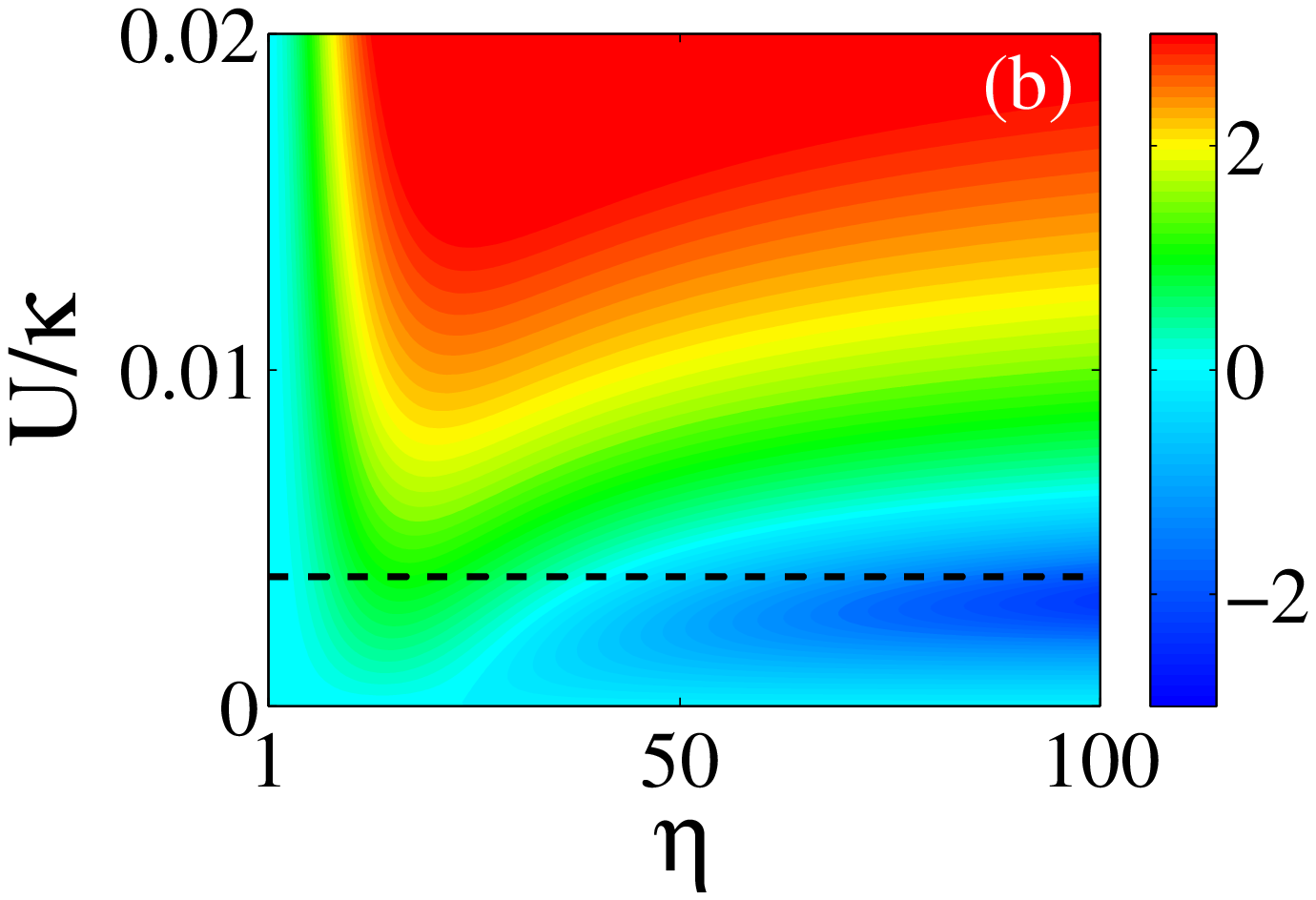}
\caption{(Color online) Logarithmic plot (of base $10$) of the equal-time
second-order correlation function $g_{a}^{\left( 2\right) }\left( 0\right)$
as a function of the strength ratio between the driving fields $\protect\eta$
and the nonlinear interaction strength $U/\kappa$ with the detuning $\Delta/\protect\kappa$ satisfying Eq.~(\ref{eq:24}) in (a) or satisfying Eq.~(\ref{eq:2}) in (b).
The dash line in (a) [(b)] corresponds to the nonlinear interaction strength $U/\kappa$ satisfying Eq.~(\ref{eq:25}) [Eq.~(\ref{eq:3})].
The parameters are the same as in Fig.~\ref{fig1}.}
\label{fig2}
\end{figure}

The equal-time second-order correlation function $g_{a}^{\left( 2\right)
}\left( 0\right) $ as a function of the strength ratio between the driving
fields $\eta$ and the nonlinear interaction strength $U/\kappa$ is shown in
Fig.~\ref{fig2} with the detuning $\Delta/\kappa$ satisfying Eq.~(\ref{eq:24}) in (a) or the detuning $\Delta/\kappa$ satisfying Eq.~(\ref{eq:2}) in (b).
Similar to the results given in Fig.~\ref{fig1}(a),
we can see from Fig.~\ref{fig2}(a) that the optimal condition for antibunching in Eq.~(\ref{eq:24}) is not applicable as $\eta$ is too large. As shown in Fig.~\ref{fig2}(b), the optimal condition Eq.~(\ref{eq:2}) becomes suitable for large enough $\eta$.

\begin{figure}[tbp]
\includegraphics[bb=36 18 395 292, width=4.15 cm, clip]{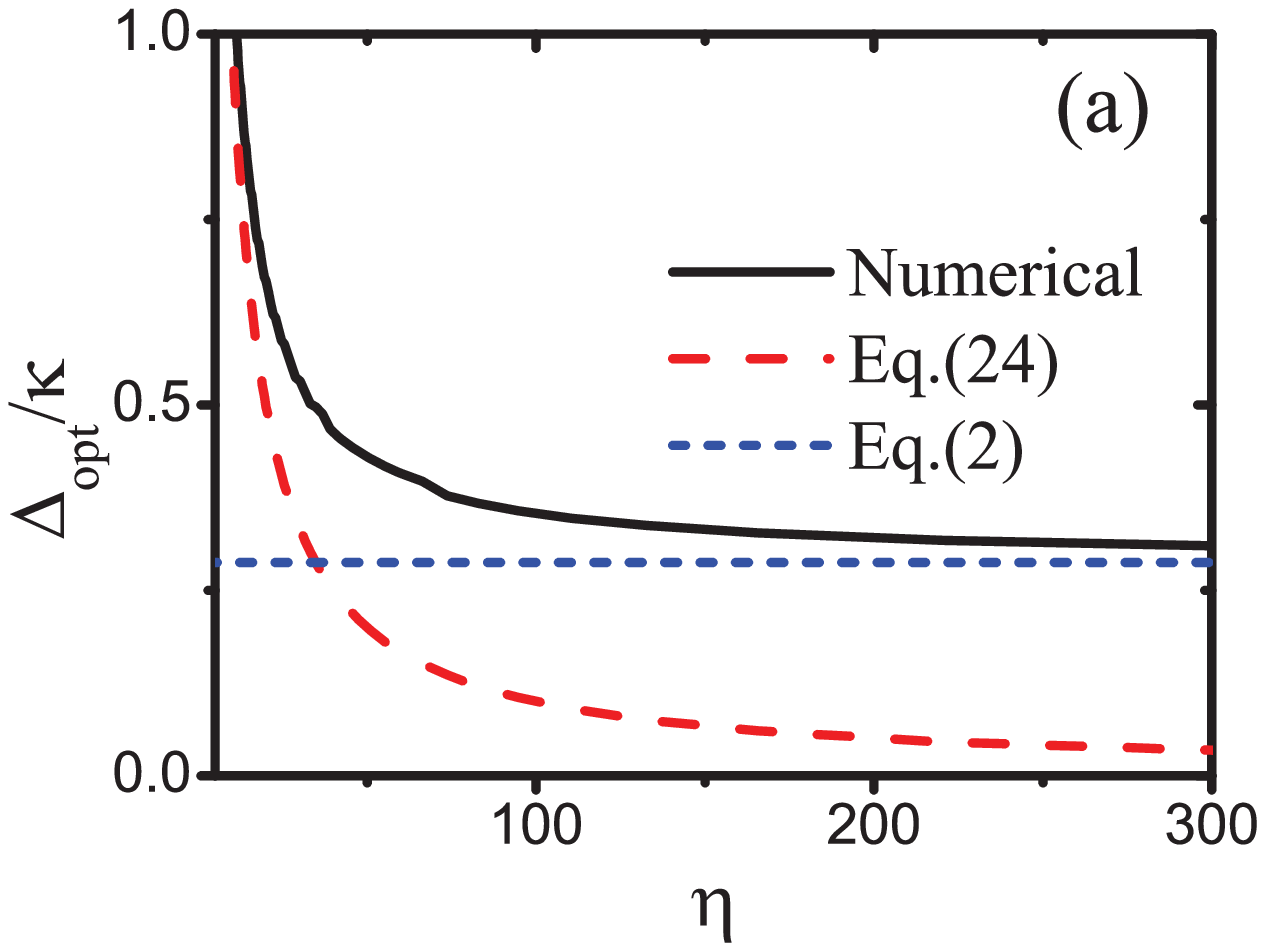} %
\includegraphics[bb=18 18 391 292, width=4.3 cm, clip]{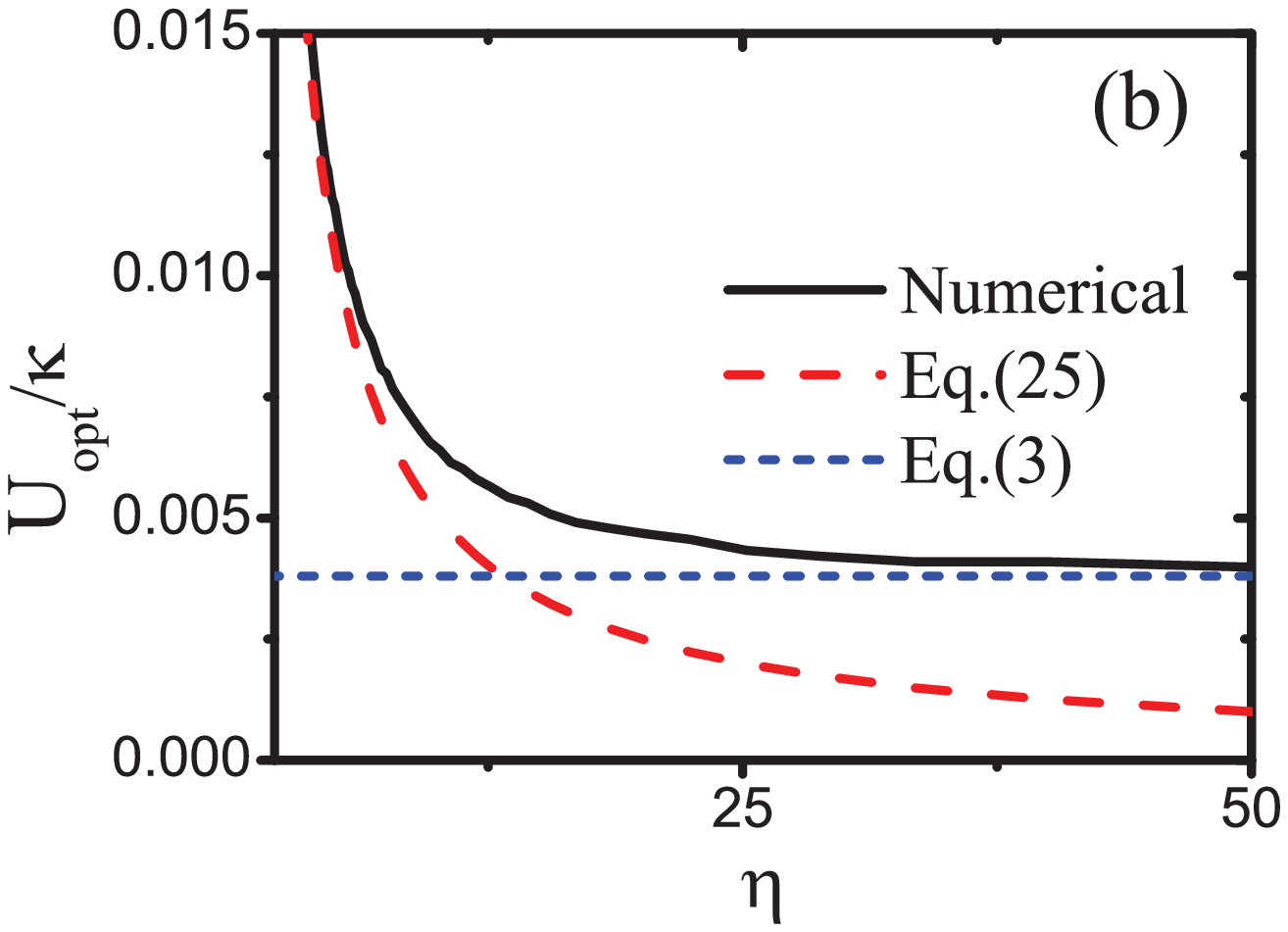}
\caption{(Color online) (a) The optimal values of the detuning $\Delta_{\rm opt}/\kappa$ for the strong antibunching effect as functions of the strength ratio between the driving fields $\eta$ according to the numerical results (black solid line),
Eq.~(\ref{eq:24}) (red dash line), and Eq.~(\ref{eq:2}) (blue short dash line).
(b) The optimal values of the nonlinear interaction strength $U_{\rm opt}/\kappa$ for strong antibunching effect as functions of $\eta$ according to the numerical results (black solid line),
Eq.~(\protect\ref{eq:25}) (red dash line), and Eq.~(\protect\ref{eq:3}) (blue short dash line).
The parameters are the same as in Fig.~\ref{fig1}.}
\label{fig3}
\end{figure}

The optimal values of the detuning $\Delta_{\rm opt}/\kappa$ and nonlinear interaction strength $U_{\rm opt}/\kappa$ for strong antibunching as functions of the strength ratio between the driving fields $\eta$ are shown in Fig.~\ref{fig3}. The black solid lines are obtained by solving Eqs.~(\ref{eq:7})-(\ref{eq:11}) in the steady state numerically and finding the optimal values of $\Delta/\kappa$ and $U/\kappa$ to minimize $g_{a}^{\left( 2\right) }\left( 0\right)$ given by Eq.~(\ref{eq:21b}); the red dash lines are plotted by Eqs.~(\ref{eq:24}) and (\ref{eq:25}); the blue short dash lines are obtained by Eq.~(\ref{eq:2}) and (\ref{eq:3}). In the regime $1<\eta < 10$,
the numerical results agree well with Eqs.~(\ref{eq:24}) and (\ref{eq:25}); as $\eta \gg 1$, the numerical results become close to the lines given by Eqs.~(\ref{eq:2}) and (\ref{eq:3}).

\begin{figure}[tbp]
\includegraphics[bb=88 278 485 551, width=4.2 cm, clip]{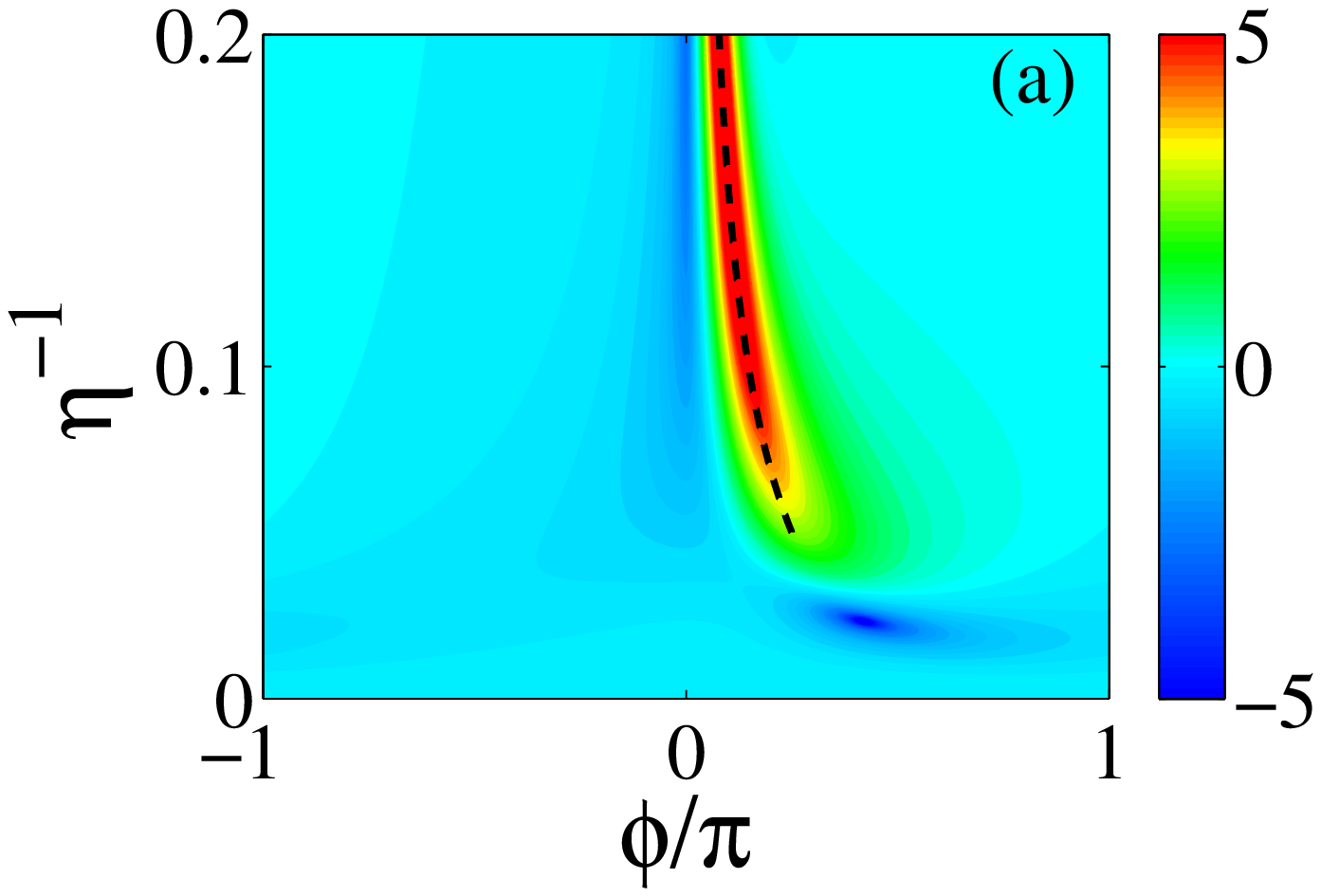} %
\includegraphics[bb=88 278 485 551, width=4.2 cm, clip]{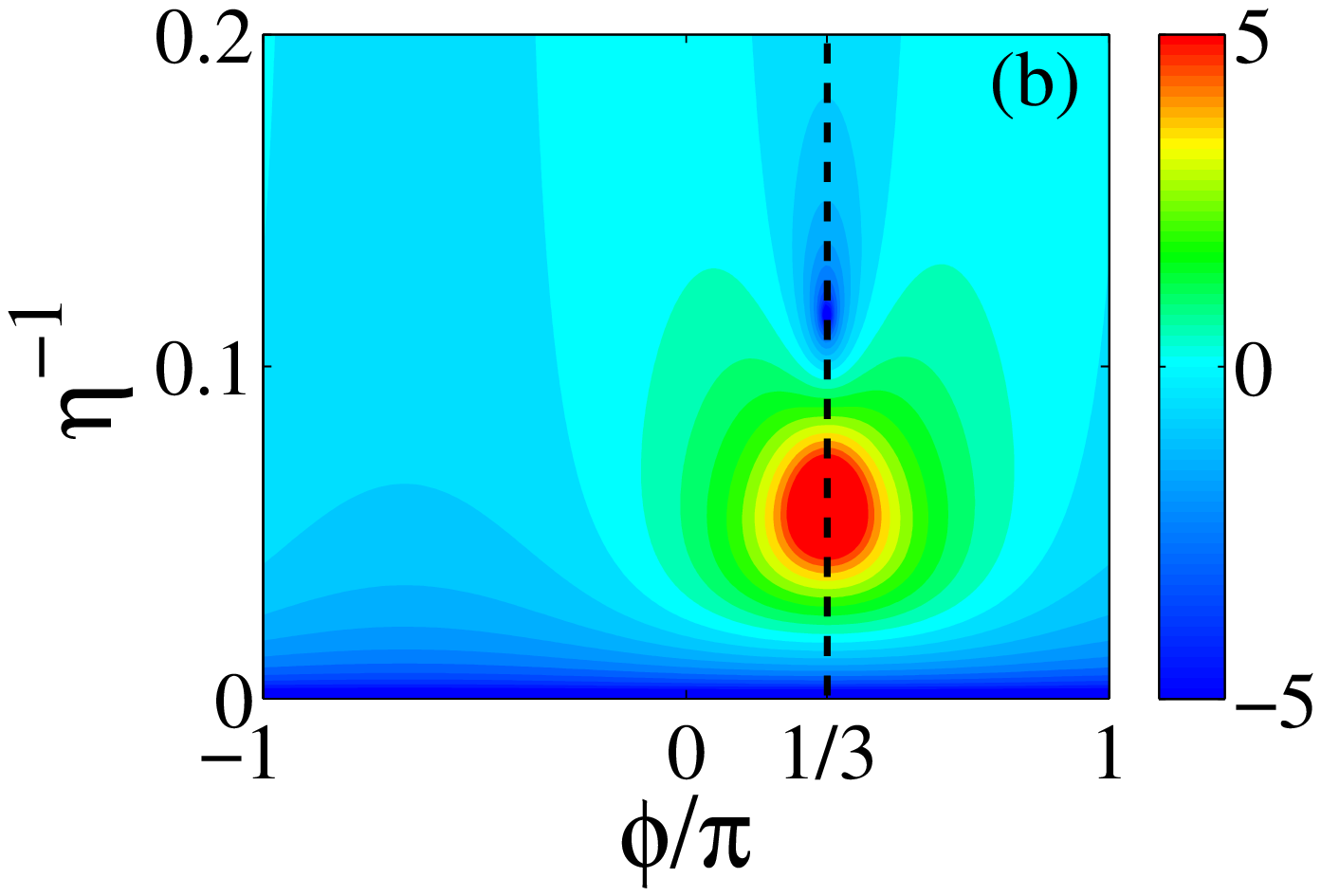} %
\includegraphics[bb=29 14 380 275, width=4.2 cm, clip]{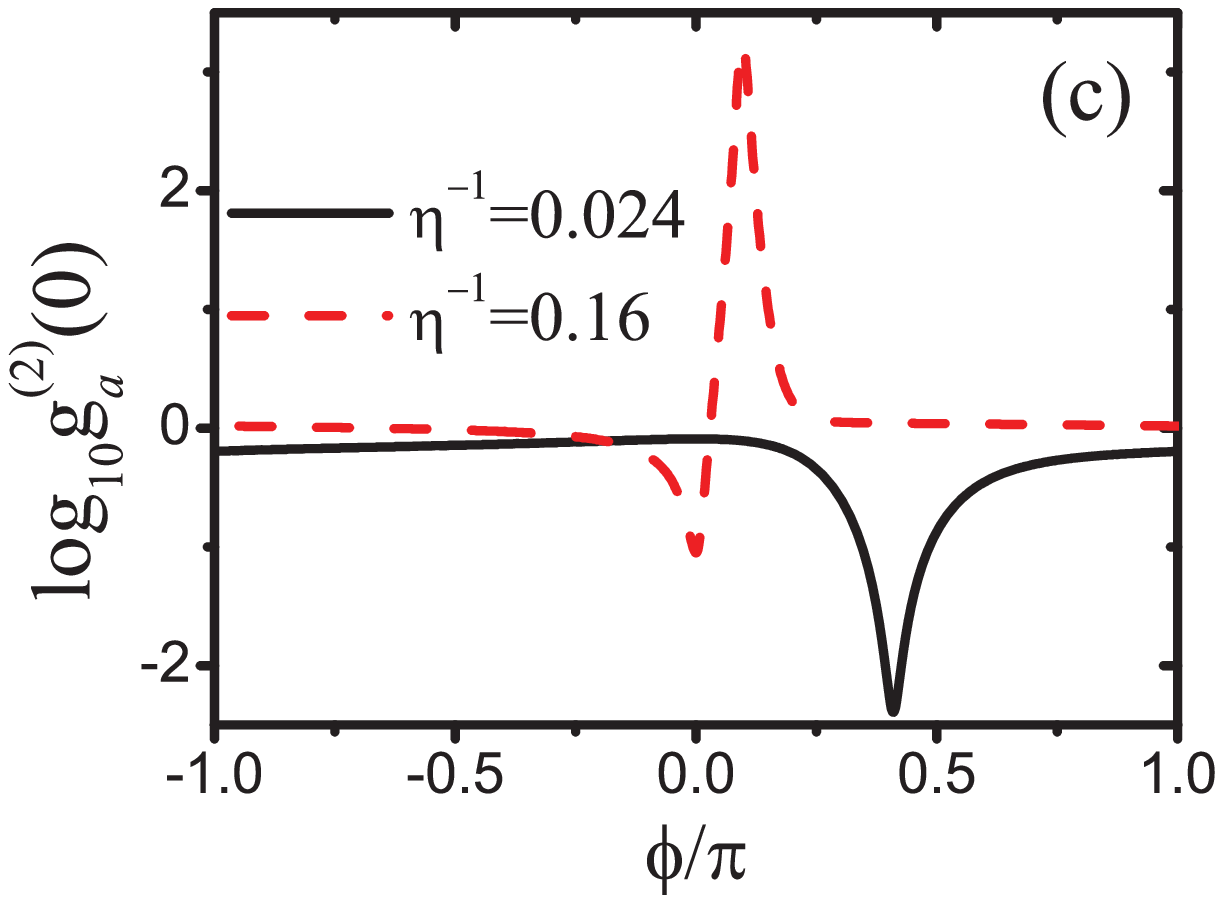} %
\includegraphics[bb=29 14 380 275, width=4.2 cm, clip]{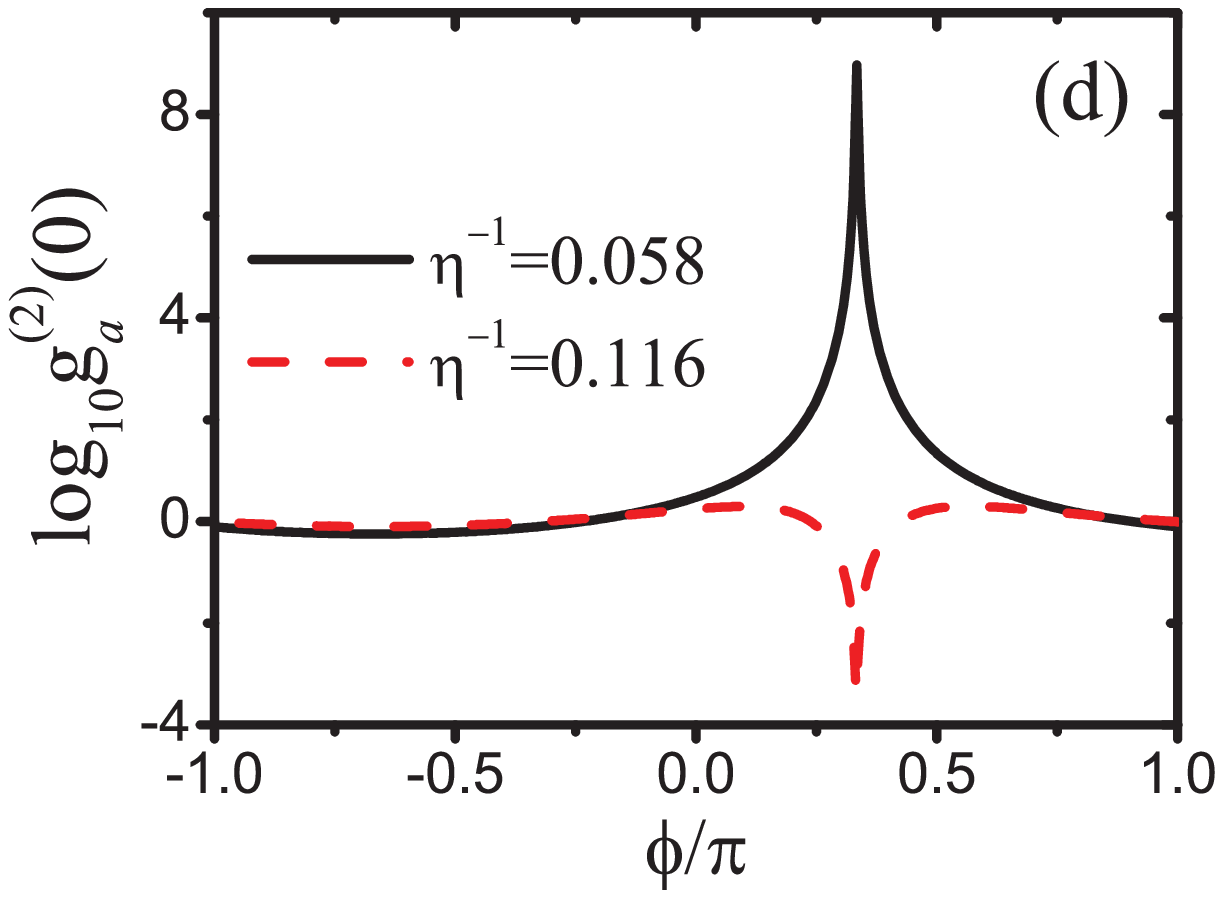}
\caption{(Color online) Logarithmic plot (of base $10$) of the equal-time
second-order correlation function $g_{a}^{\left( 2\right) }\left( 0\right)$
as a function of the relative phase $\protect\phi$ and the strength
ratio $\protect\eta^{-1}$ between the driving fields with the
parameters $\Delta/\kappa$ and $U/\kappa$ given by Eqs.~(\protect\ref{eq:24})-(\protect\ref{eq:25}) in (a)
or the parameters $\Delta/\kappa$ and $U/\kappa$ given by Eqs.~(\protect\ref{eq:2})-(\protect\ref%
{eq:3}) in (b). $g_{a}^{\left( 2\right)}\left( 0\right)$ as functions of the
relative phase $\protect\phi$ for different values of the strength
ratio $\eta$ is shown (c) and (d), i.e. a few cuts taken from the color
plot (a) and (b), respectively. The coupling constant $J =10\kappa$.
}
\label{fig4}
\end{figure}

Besides the strength ratio $\eta$, the relative phase $\phi$
between the driving fields is another controllable parameter in
experiments. Two-dimensional plots of the equal-time second-order
correlation function $g_{a}^{\left( 2\right) }\left( 0\right)$ as a function
of the relative phase $\phi$ and the strength ratio $\eta^{-1}$
between the driving fields is shown in Fig.~\ref{fig4} with the
parameters $\Delta/\kappa$ and $U/\kappa$ given by Eqs.~(\ref{eq:24}) and (\ref{eq:25}) in Fig.~\ref{fig4}(a) or given by Eqs.~(\ref{eq:2}) and (\ref{eq:3}) in Fig.~\ref{fig4}(b). From these figures,
we can see that: (i) the photon statistic properties are dependent on both the
relative phase and strength ratio between the driving fields; (ii) there
are not only strong antibunching but also strong bunching effect for optimal
relative phase and strength ratio between the driving fields. The
equal-time second-order correlation function $g_{a}^{\left( 2\right)}\left(
0\right)$ as functions of the relative phase $\phi$ for different
strength ratio between the driving fields $\eta^{-1}$ are shown in Figs.~\ref{fig4}(c) and (d), which are a few cuts taken from Figs.~\ref{fig4}(a) and (b), respectively. From Figs.~\ref{fig4}(c) and (d), we find that the photon
statistic properties can be controlled by tuning the relative phase $\phi$ in different ways for the strength ratio $\eta^{-1}$ taking different values. As shown in
Fig.~\ref{fig4}(c), as $\eta^{-1}=0.024$, there is a regime for strong
antibunching around $\phi =0.41 \pi$ [see the black solid line in Fig.~\ref{fig4}(c)]; if $\eta^{-1}=0.16$, the photons exhibit antibunching as $\phi=0$ but exhibit bunching as $\phi=0.096 \pi$ [see the red dash line in Fig.~\ref{fig4}(c)]. In Fig.~\ref{fig4}(d), for $\phi=\pi/3$, the photons exhibit strong bunching as $\eta^{-1}=0.058$ (see the corresponding black solid line) but exhibit strong antibunching as $\eta^{-1}=0.116$ (see the corresponding red dash line).

\begin{figure}[tbp]
\includegraphics[bb=20 5 387 281, width=4.2 cm, clip]{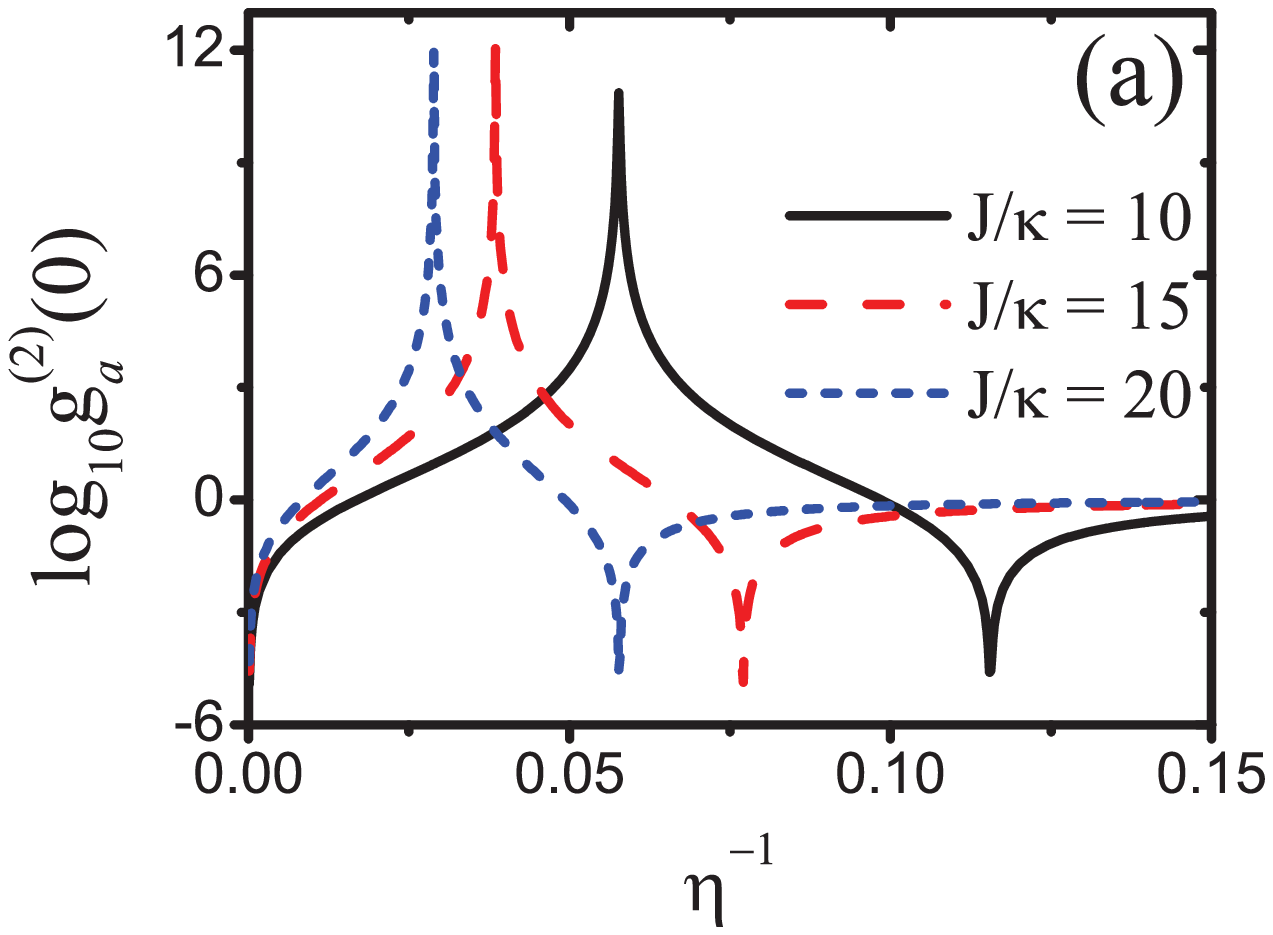} %
\includegraphics[bb=20 4 387 278, width=4.2 cm, clip]{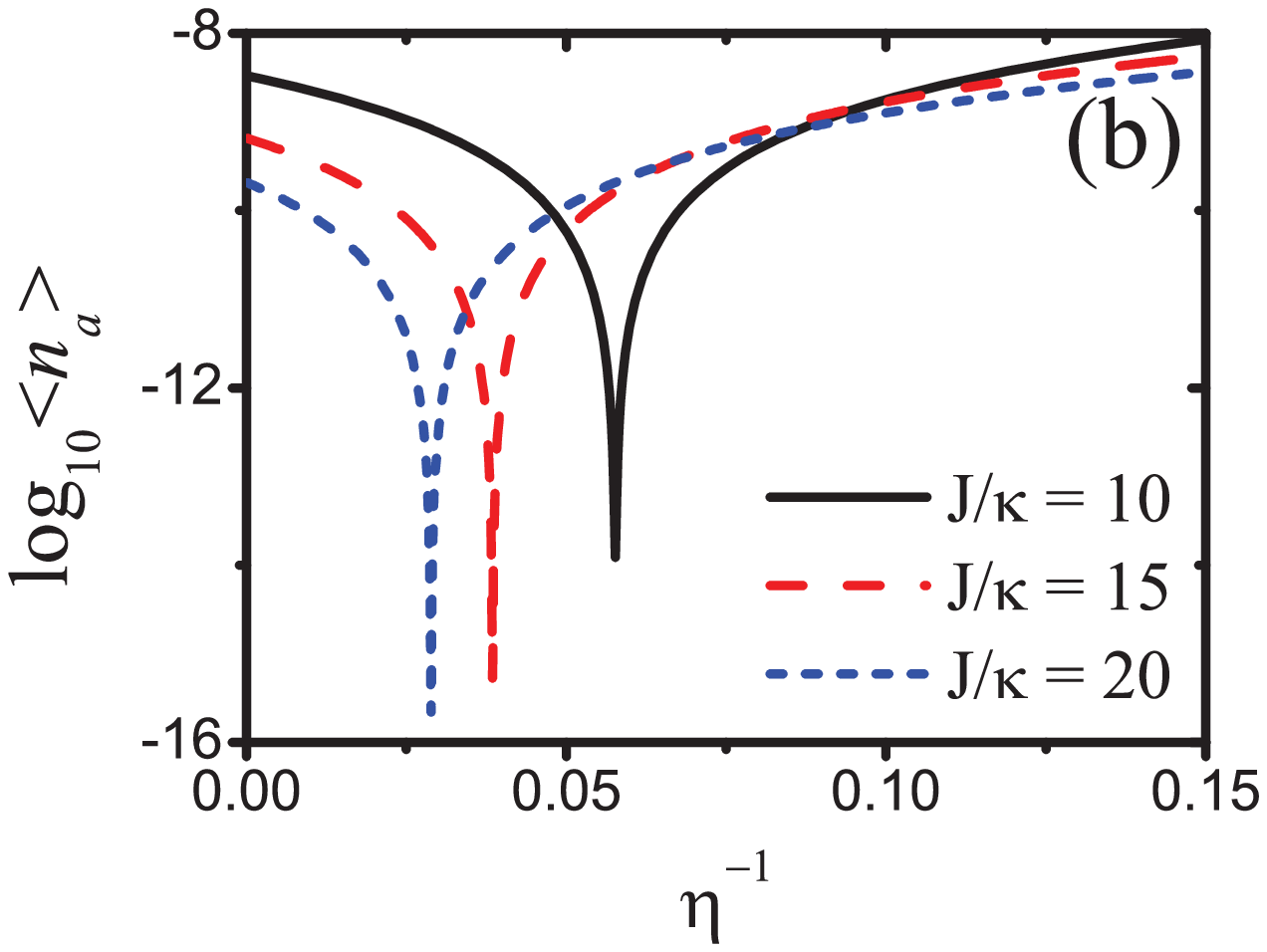}
\caption{(Color online) Logarithmic plot (of base $10$) of (a) the
equal-time second-order correlation function $g_{a}^{\left( 2\right) }\left(
0\right)$ and (b) the mean photon number $\langle n_{a} \rangle$ as functions of
the strength ratio $\protect\eta^{-1}$ for different values of the coupling
constant $J/\protect\kappa$ with the parameters $\Delta/\kappa$ and $U/\kappa$ given by Eqs.~(\ref{eq:2})-(\ref{eq:3}), and the relative phase $\phi = \pi/3$.}
\label{fig5}
\end{figure}

It is worth analyzing the seasons for arising strong bunching effect in Fig.~%
\ref{fig4}. We take the strong bunching regime around the point ($\phi=\pi/3$, $\eta^{-1}=0.058$) in Fig.~\ref{fig4}(b) for an example. Using the optimal parameters satisfying Eqs.~(\ref{eq:2}) and (\ref{eq:3}), the logarithmic plot (of base $10$) of $g_{a}^{\left( 2\right) }\left( 0\right) $ [Fig.~\ref{fig5}(a)] and the mean photon number $\langle
n_{a}\rangle $ [Fig.~\ref{fig5}(b)] as functions of the strength ratio $\eta ^{-1}$ for different values of the coupling constant $J/\kappa $ are shown in Fig.~\ref{fig5}. We can see that the bunching regime is accompanied by a strong
suppression of the photon number, and the similar phenomenon has been reported
in Ref.~\cite{KyriienkoarX14}. In other words, in the condition for strong
bunching effect, the probability of generating photon pair increases, while the probability for single-photon emission decreases. This process is called photon-induced tunnelling~\cite{DayanSci08}. The photon-induced tunnelling can be applied to controlled photonic quantum gates~\cite{XuPRA13,KubanekPRL08,KochPRL11}.

Physically, the strong suppression of the photon number and strong bunching effect shown here originate from the destructive interference between the two paths for generating photons in cavity mode A: (i) the direct photon excitation in mode A and (ii) exciting photons in mode B then tunneling into mode A. In order to show the origin of the quantum interference in detail, we will treat this problem mathematically. In the weak driving condition $\varepsilon _{a,b}\ll \kappa $, we have $\left\langle n_{a}\right\rangle
\approx |C_{10}|^{2}$. From Eq.~(\ref{eq:17}), the condition for $C_{10}=0$ is given by%
\begin{equation}
J-\eta e^{i\phi }\left( \Delta -\frac{i\kappa }{2}\right) =0.  \label{eq:26}
\end{equation}%
With the parameters $\Delta/\kappa$ and $U/\kappa$ given by Eqs.~(\ref{eq:2}) and (\ref{eq:3}), $%
\left\langle n_{a}\right\rangle \approx 0$ and strong bunching effect are
obtained around the point
\begin{equation}
\phi =\frac{\pi }{3}, \quad \eta ^{-1}=\frac{\kappa }{\sqrt{3}J}.
\end{equation}%
These agree well with the results shown in Fig.~\ref{fig5}. Similarly,
for the optimal parameters given by Eqs.~(\ref{eq:24}) and (\ref{eq:25}), when $\eta \kappa/2J \ll 1$, the strong bunching effect appears along the black dash curve for
\begin{equation}
\phi = \tan^{-1}\left( \frac{\eta \kappa}{2J} \right),
\end{equation}
in Fig.~\ref{fig4}(a).

\section{Conclusions}

In summary, we have studied the photon statistics in a nonlinear photonic molecule for
both the two cavity modes being driven coherently. By analytical and numerical
methods, we find that the optimal parameters for strong photon antibunching or
bunching effects are related to the coupling constant between the cavity modes, the
strength ratio and the relative phase between the two driving fields. Thus
we can control the statistic properties of the photons by tuning these parameters.
Future applications for nonlinear photonic molecules with both the two cavity modes
being driven coherently include the tunable
single-photon sources and the controlled photonic quantum gates.

\vskip 2pc \leftline{\bf Acknowledgement}

We thank Q. Zheng and Y. Yao for fruitful discussions. This work is
supported by the Postdoctoral Science Foundation of China (under Grant No. 2014M550019), the NSFC (under Grant No. 11174027), and the National 973 program (under Grant No. 2012CB922104 and No. 2014CB921402).

\bibliographystyle{apsrev}
\bibliography{ref}

\end{document}